\documentstyle[12pt,epsf,epsfig]{article}
\newcount\timecount
\newcount\hours \newcount\minutes  \newcount\temp \newcount\pmhours

\hours = \time
\divide\hours by 60
\temp = \hours
\multiply\temp by 60
\minutes = \time
\advance\minutes by -\temp
\def\hour{\the\hours}
\def\minute{\ifnum\minutes<10 0\the\minutes
            \else\the\minutes\fi}
\def\clock{
\ifnum\hours=0 12:\minute\ AM
\else\ifnum\hours<12 \hour:\minute\ AM
      \else\ifnum\hours=12 12:\minute\ PM
            \else\ifnum\hours>12
                 \pmhours=\hours
                 \advance\pmhours by -12
                 \the\pmhours:\minute\ PM
                 \fi
            \fi
      \fi
\fi
}

\def\monthname{\relax\ifcase\month 0/\or January\or February\or
   March\or April\or May\or June\or July\or August\or September\or
   October\or November\or December\else\number\month/\fi}

\def\bold#1{\setbox0=\hbox{$#1$}%
     \kern-.025em\copy0\kern-\wd0
     \kern.05em\copy0\kern-\wd0
     \kern-.025em\raise.0433em\box0 }

\def\gappeq{\mathrel{\rlap {\raise.5ex\hbox{$>$}}
{\lower.5ex\hbox{$\sim$}}}}

\def\lappeq{\mathrel{\rlap{\raise.5ex\hbox{$<$}}
{\lower.5ex\hbox{$\sim$}}}}

\def\ga{\mathrel{\raise.3ex\hbox{$>$\kern-.75em\lower1ex\hbox{$\sim$}}}}
\def\la{\mathrel{\raise.3ex\hbox{$<$\kern-.75em\lower1ex\hbox{$\sim$}}}}
\def\gev{{\rm \, Ge\kern-0.125em V}}
\def\tev{{\rm \, Te\kern-0.125em V}}
\def\beq{\begin{equation}}
\def\eeq{\end{equation}}

\def\ohsq{\Omega_{\chi} h^2}

\def\m12{m_{1\!/2}}

\begin{document}
\begin{titlepage}
\pagestyle{empty}
\baselineskip=21pt
\rightline{hep-ph/0105004}
\rightline{CERN--TH/2001-111}
\rightline{UMN--TH--2002/01, TPI--MINN--01/16}
\vskip 0.35in
\begin{center}
{\large{\bf How Finely Tuned is Supersymmetric Dark Matter?
}}
\end{center}
\begin{center}
\vskip 0.05in
{{\bf John Ellis}$^1$ and
{\bf Keith A.~Olive}$^{1,2}$
\vskip 0.05in
{\it
$^1${TH Division, CERN, Geneva, Switzerland}\\
$^2${Theoretical Physics Institute, School of Physics and Astronomy,\\
University of Minnesota, Minneapolis, MN 55455, USA}\\
}}
\vskip 0.35in
{\bf Abstract}
\end{center}
\baselineskip=18pt \noindent

We introduce a quantification of the question in the title: the
logarithmic sensitivity of the relic neutralino density $\ohsq$ to
variations in input parameters such as the supersymmetric mass scales
$m_0, m_{1/2}$ and $A_0$, $\tan \beta$ and the top and bottom quark
masses. In generic domains of the CMSSM parameter space with a relic
density in the preferred range $0.1 \le \ohsq \le 0.3$, the sensitivities
to all these parameters are moderate, so an interesting amount of
supersymetric dark matter is a natural and robust prediction. Within these
domains, the accuracy in measuring the CMSSM and other input parameters at
the LHC may enable the relic density to be predicted quite precisely. 
However, in the coannihilation regions, this might require more
information on the supersymetric spectrum than the LHC is able to provide. 
There are also exceptional domains, such as those where direct-channel
pole annihilation dominates, and in the `focus-point' region, where the
logarithmic sensitivity to the input parameters is greatly increased, and
it would be more difficult to predict $\ohsq$ accurately. 

\vfill
\vskip 0.15in
\leftline{CERN--TH/2001-111}
\leftline{May 2001}
\end{titlepage}
\baselineskip=18pt

The annihilations of stable particles weighing $\lappeq 100$~TeV that were
once in thermal equilibrium in the early Universe are able to produce a
relic density comparable to the critical density. In particular,
weakly-interacting stable particles weighing $\sim 1$~TeV may well have a
cosmological density in the preferred range, if they were formerly in
thermal equilibrium.  An example is provided by the lightest
supersymmetric particle, assumed to be the lightest neutralino $\chi$,
which is expected to be stable in models with conserved $R$
parity~\cite{Fayet}.  For example, it is often remarked that
supersymmetric dark matter `naturally' has a relic density $\ohsq$ in the
range $0.1 \le \ohsq \le 0.3$ preferred by astrophysics and
cosmology~\cite{EHNOS}.

The TeV mass scale for supersymmetry is motivated independently by the
hierarchy problem: how to make the small electroweak scale $m_W \ll m_P
\sim 10^{19}$~GeV `natural', without the need to fine-tune parameters at
each order in perturbation theory~\cite{hierarchy}. This is possible if
the supersymmetric partners of the Standard Model particles weigh $\lappeq
1$~TeV, but the amount of fine-tuning of supersymmetric parameters
required to obtain the electroweak scale increases rapidly for sparticle
masses $\gg 1$~TeV. In an attempt to quantify this, it was
proposed~\cite{EENZ,BG} to consider the logarithmic sensitivities of the
electroweak scale to the supersymmetric model parameters $a_i$:
\begin{equation}
\Delta_i \equiv {a_i \over m_W} {\partial m_W \over \partial a_i}.
\label{EENZ}
\end{equation}
In the constrained MSSM (CMSSM) with universal soft supersymmetry-breaking
parameters, the $a_i$ include the common scalar mass $m_0$, the common
gaugino mass $m_{1/2}$, the common trilinear parameter $A_0$ at the GUT
scale and the ratio of Higgs vev's, $\tan \beta$, with the Higgs mixing
parameter $\mu$ being determined (up to a sign) by the electroweak vacuum
conditions. The measure (\ref{EENZ}) has been used, for
example, to quantify the fine-tuning price imposed by the absence of
sparticles at LEP~\cite{CEP}.  The point has also been made that
supersymmetric models with $0.1 \le \ohsq \le 0.3$ tend to have small
values of $\Delta \equiv {\rm Max}_i \Delta_i$~\cite{CEOP}, establishing a
link between (the absence of) hierarchical fine-tuning and good cosmology. 

In this paper, we propose analogous measures of sensitivity to quantify
the fine-tuning needed to obtain in the CMSSM a relic density
$\ohsq$ in the range preferred by cosmology: 
\begin{equation}
\Delta^\Omega_i \equiv {a_i \over \Omega_\chi} {\partial \Omega_\chi \over
\partial a_i}.
\label{EO}    
\end{equation}
The input parameters $a_i$ now include, along with the CMSSM parameters
introduced above, the top- and bottom-quark masses, Standard Model
parameters which are not so well known, and whose current
uncertainties have important impacts on calculations of $\ohsq$. We also
explore
the accuracy to which measurements of the CMSSM parameters at the LHC might
enable $\ohsq$ to be calculated~\cite{Drees}. 

In generic regions of the CMSSM parameter space, we find that the
overall sensitivity
\begin{equation}
\Delta^\Omega \equiv \sqrt{\Sigma_i (\Delta^\Omega_i)^2}
\label{sensitivity}
\end{equation}
is relatively small: $\Delta^\Omega \lappeq 10$, implying that
measurements of the input parameters at the 10 [1] \% level will enable
$\ohsq$ to be calculated to within a factor ${\cal O} (2)$ [${\cal O} (10
\%)$]. The sensitivity $\Delta^\Omega$ is somewhat enhanced in the $\chi -
{\tilde \ell}$ coannihilation region~\cite{EFOSi,glp}, and here an
accurate
calculation of the relic density might not be possible with LHC measurements
of the CMSSM parameters alone. There are also exceptional regions where the
sensitivity of $\Delta^\Omega$ is greatly enhanced, notably at large $\tan
\beta$ where there are `funnels' in CMSSM parameter space due to rapid $\chi
\chi
\rightarrow H, A$ annihilation~\cite{EFGOSi}, and in
the `focus-point' region~\cite{FM}, where $\Delta^\Omega$ may rise to several
hundred.  In the focus-point region, there is extreme
sensitivity to $m_t$:  even if $m_t$ is measured at the 1 \% level, $\ohsq$
may be uncertain by a large factor for any specific set of CMSSM
parameters. 

We start by outlining our procedure~\cite{EFGOSi} for calculating the
neutralino relic density $\ohsq$ and its sensitivity to the CMSSM
parameters. As already mentioned, we consider as independent parameters
the universal soft mass terms $m_0, m_{1/2}$, the trilinear soft
supersymmetry-breaking parameter $A_0$, and $\tan \beta$. We also assume
unification of the gauge couplings at the GUT scale as an input into the
renormalization-group calculations of the CMSSM parameters at the
electroweak scale. The top- and bottom-quark masses are potentially
important for the relic density calculations, particularly at large $\tan
\beta$, and are relatively poorly known, so we also track the sensitivity of
$\ohsq$ to their values. As defaults, we choose the running bottom-quark mass
$m_b(m_b)^{\overline {MS}}_{SM} = 4.25$~GeV~\cite{mb} and the top-quark pole
mass $m_t = 175$~GeV. However, for our calculations in the `focus-point'
region~\cite{FM} we use $m_t = 171$~GeV. This choice of $m_t$ allows us to
display the focus-point region at values of $m_0$ between 1 and 2 TeV,
for ease of comparison with~\cite{FM}. If we had chosen $m_t = 175$ GeV,
our calculations would have located the focus-point region 
between 2 and 3 TeV.

More details of our code to evaluate $\ohsq$ are given in~\cite{EFGOSi}
and references therein, so here we note just a few relevant aspects. 
Calculations at small-to-moderate $\tan \beta \lappeq 25$ have no novel
features, though we do recall the importance of including $\chi - {\tilde
\ell}$ coannihilation processes at large $m_{1/2}$. As discussed
in~\cite{EFGOSi}, several new coannihilation processes and diagrams become
relevant at larger values of $\tan \beta$, which are included here. Also
important at large $\tan \beta$ are direct-channel annihilation processes: 
$\chi \chi \rightarrow H,A$, where $H,A$ are the heavier neutral Higgs
bosons in the CMSSM.  Their treatment requires going beyond~\cite{directpoles}
the non-relativistic partial-wave expansion that is adequate elsewhere. 

In order to calculate the sensitivities $\Delta^\Omega_i$ (\ref{EO}), we
first define a grid in the $(m_{1/2}, m_0)$ plane for fixed $A_0, \tan
\beta, m_t$ and $m_b$, on which we compute the values of $\ohsq$. We then
compute the differences in $\ohsq$ generated by small ($\lappeq
1$~\%) changes in each of
$m_{1/2}, m_0, A_0, \tan \beta, m_t$ and $m_b$ individually. We then use
these small finite differences to calculate the various sensitivities
$\Delta^\Omega_i$ (\ref{EO}) and hence the overall sensitivity
$\Delta^\Omega$ (\ref{sensitivity}). Thus, obtaining our results is quite
computation-intensive, with each of the $(m_{1/2}, m_0)$ planes that we
show below requiring several times more CPU time than the calculations of
$\ohsq$ shown previously~\cite{EFGOSi}. For this reason, we have not
increased the grid resolution sufficiently to clarify all the fluctuations
(or small effects?) that we find in this analysis. 

\begin{figure}
\vspace*{-0.75in}
\hspace*{-.20in}
\begin{minipage}{8in}
\epsfig{file=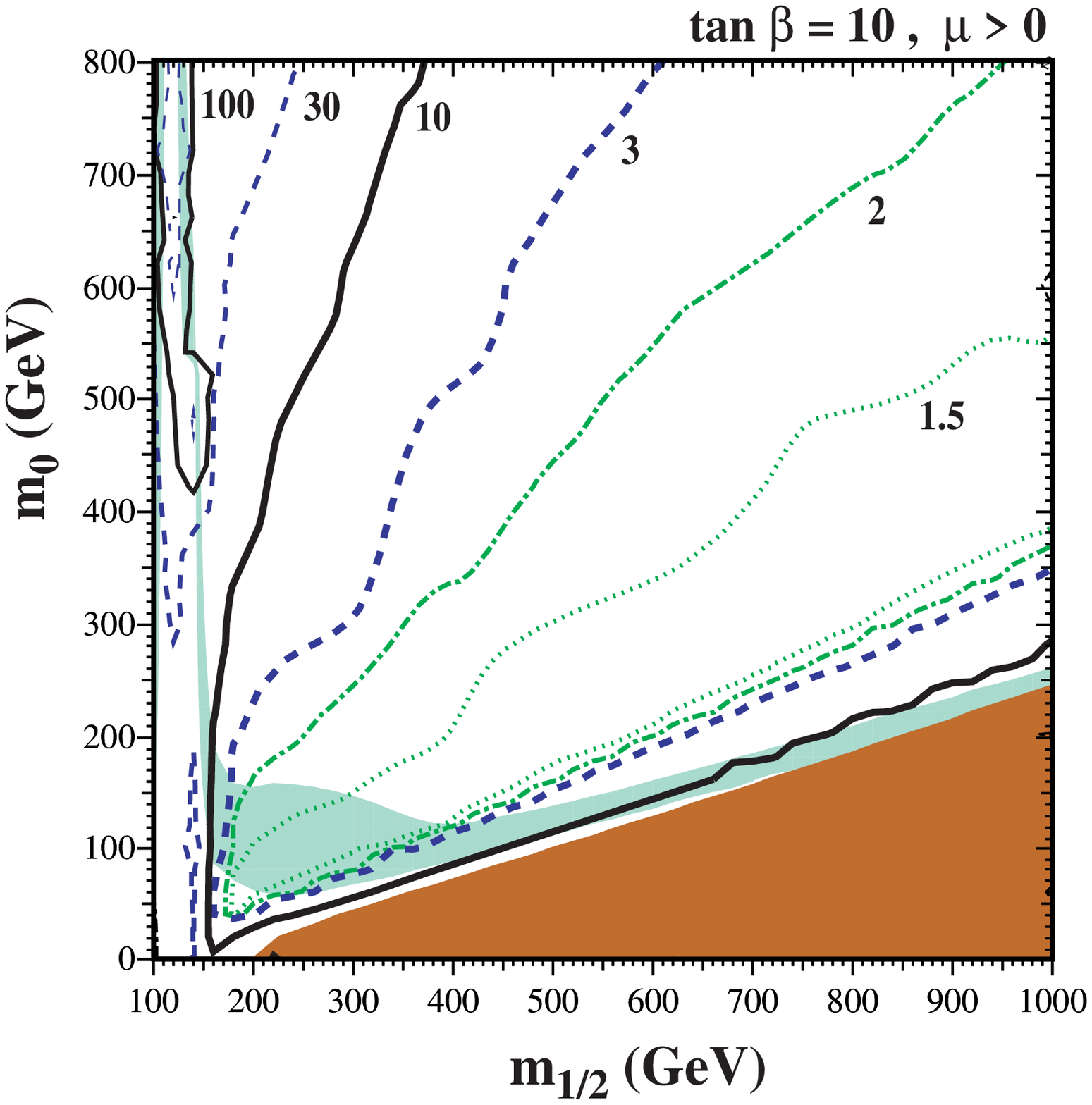,height=3.5in}
\epsfig{file=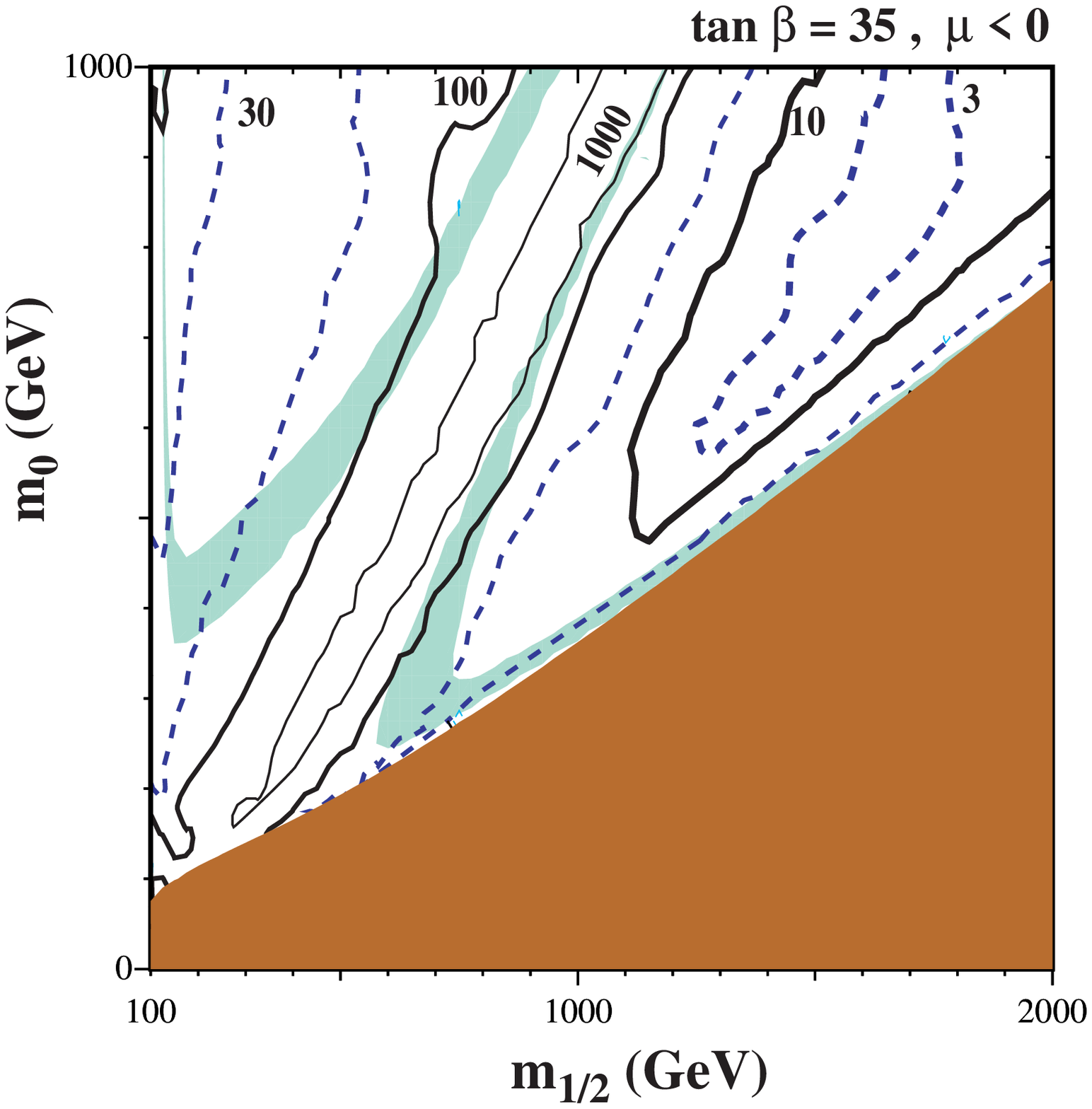,height=3.5in} \hfill
\end{minipage}
\hspace*{-.20in}
\begin{minipage}{8in}
\epsfig{file=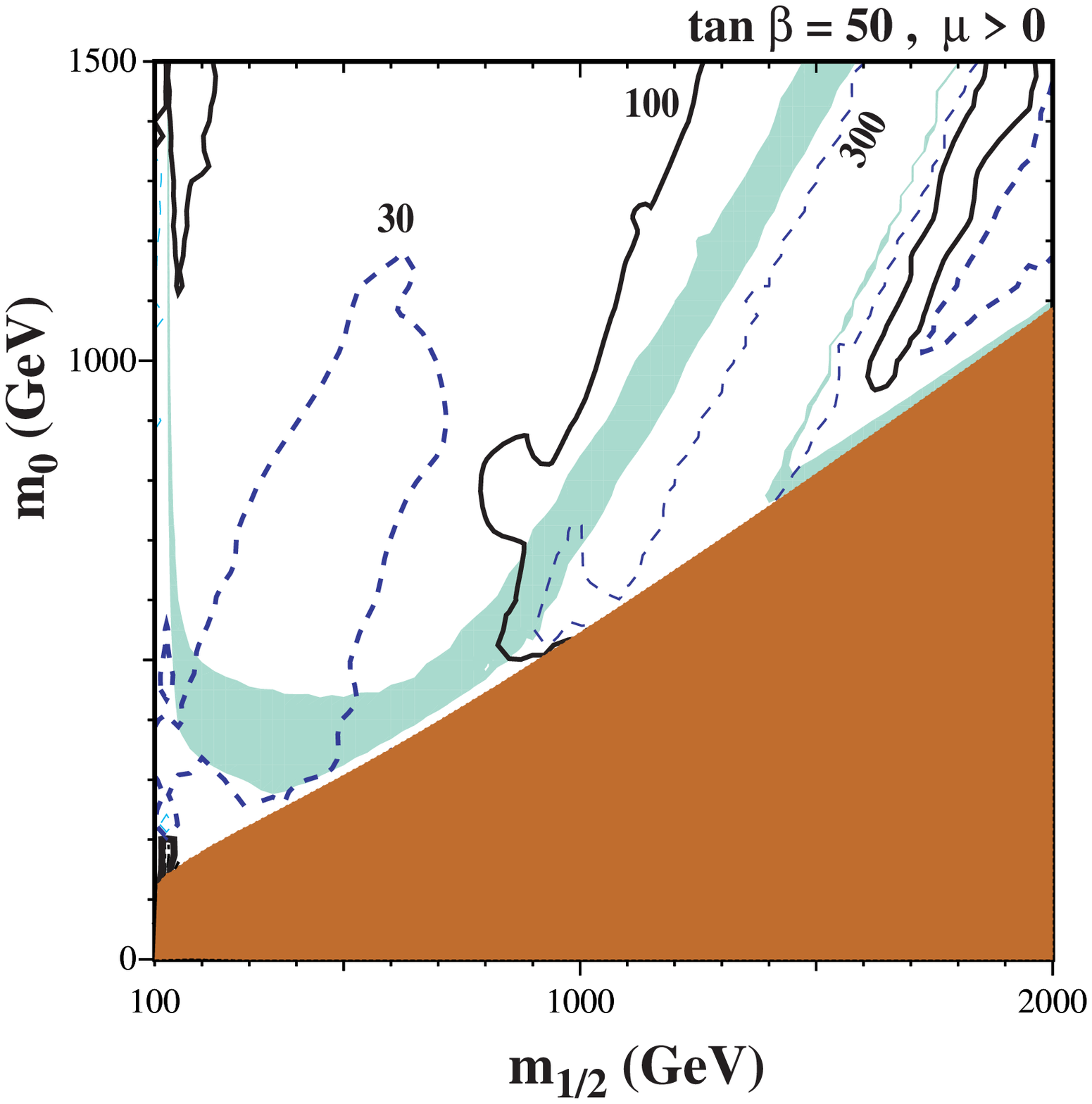,height=3.5in}
\hspace*{-0.25in}
\epsfig{file=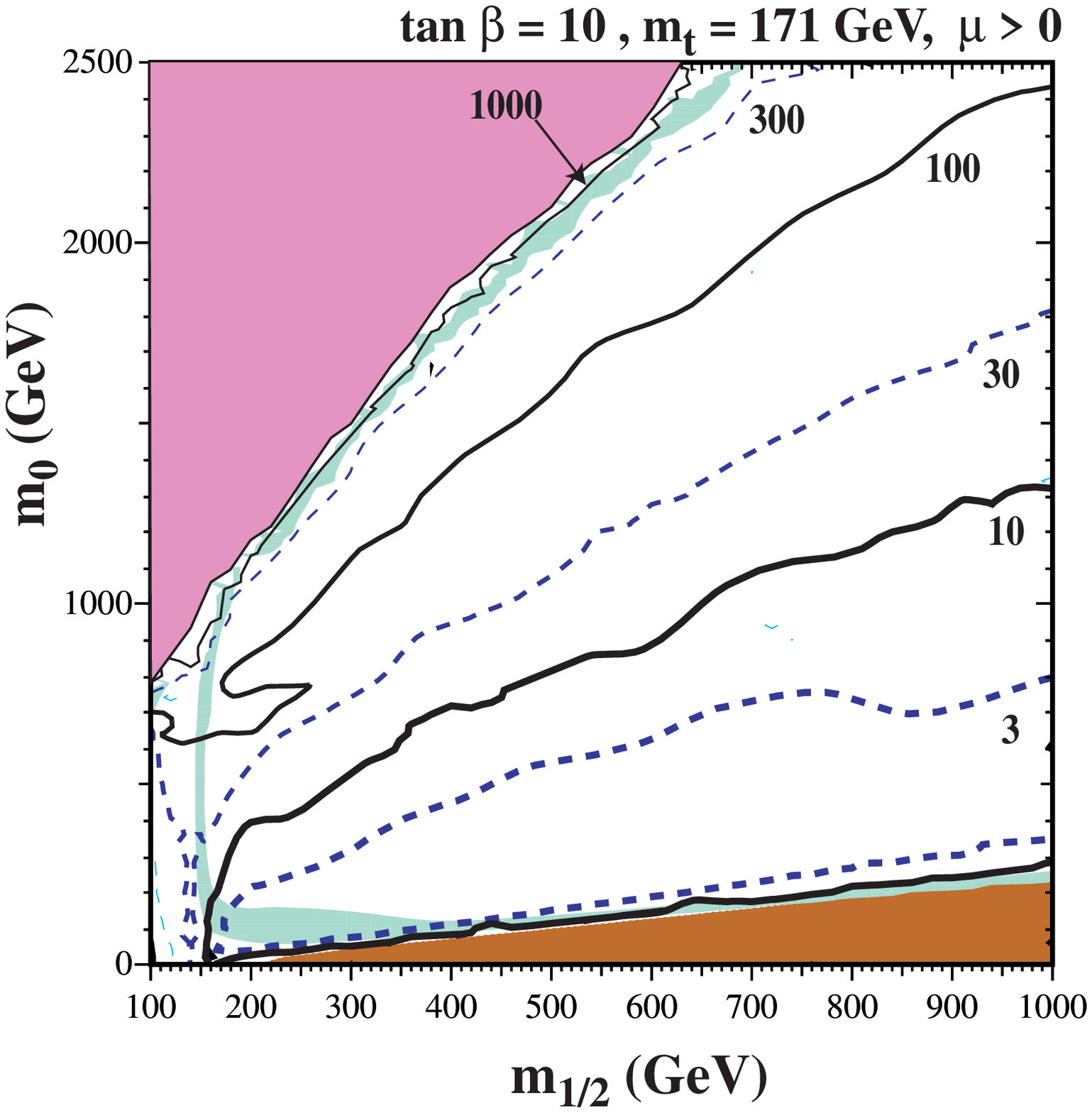,height=3.5in} \hfill
\end{minipage}
\caption{\label{fig:overall}
{\it Contours of the total sensitivity $\Delta^\Omega$ in the
$(m_{1/2}, m_0)$ planes for (a) $\tan \beta = 10, \mu > 0, m_t =
175$~GeV, (b) $\tan \beta = 35, \mu < 0, m_t = 175$~GeV, (c)
$\tan \beta = 50, \mu > 0, m_t = 175$~GeV, and (d) $\tan \beta =
10, \mu > 0, m_t = 171$~GeV, all for $A_0 = 0$. The light (turquoise)
shaded areas are the cosmologically preferred regions with
\protect\mbox{$0.1\leq\ohsq\leq 0.3$}. In the dark (brick red) shaded
regions, the LSP is the charged ${\tilde \tau}_1$, so this region is
excluded. In panel (d), the medium shaded (mauve) region is excluded by
the electroweak vacuum conditions. The curves are contours of constant
$\Delta^\Omega$, as described in the text. }}
\end{figure}

Fig.~\ref{fig:overall} displays the overall sensitivity $\Delta^\Omega$
(\ref{sensitivity}) in the $(m_{1/2}, m_0)$ planes for four representative
choices of the other CMSSM parameters. In each of these planes, the
regions with relic density in the preferred range $0.1 \le \ohsq \le 0.3$
are indicated by lighter shading, and the disallowed regions where the
lightest supersymmetric particle is the ${\tilde \tau}_1$, rather than the
lightest neutralino $\chi$, are shown by darker shading.  In these
figures, we show contours of constant values of the fine-tuning parmeter
$\Delta^\Omega$.  Contours of $\Delta^\Omega = 3, 30$, and 300 are shown
by dashed (blue)  curves of decreasing thickness.  Contours of
$\Delta^\Omega = 10, 100$, and 1000 are shown by solid (black) curves also
of decreasing thickness. In Figure ~\ref{fig:overall}a, we show the two
additional contours $\Delta^\Omega = 1.5$ and 2, as dotted and dot-dashed
curves respectively. 

Consider first panel (a), for $\tan \beta = 10, A_0 = 0, \mu > 0, m_t =
175$~GeV and $m_b(m_b)^{\overline {MS}} = 4.25$~GeV. We see that, in a
`generic' domain of the $(m_{1/2}, m_0)$ plane for moderate values of
$m_0 / m_{1/2}$ in the approximate range 1/3 to 2, the overall
sensitivity is also moderate: $\Delta^\Omega < 10$.  Indeed, there is a
substantial domain of this $(m_{1/2}, m_0)$ plane where the sensitivity
parameter $\Delta^\Omega < 3$. Therefore, at least in this domain of
parameter space, {\it supersymmetric dark matter does not require fine
tuning.} We also note that the CMSSM value of $g_\mu - 2$ is in good
agreement~\cite{ENO} with the data~\cite{g-2} in this `generic' domain at
moderate $m_0 / m_{1/2}$, as is the rate for $b \rightarrow s
\gamma$~\cite{bsg}.

Moreover, the small magnitude of $\Delta^\Omega$ {\it suggests that one
might hope, with a \% accuracy in the CMSSM parameters, to aim at a 10 \%
accuracy in calculating $\ohsq$.} In this connection, we note that the
preferred range $0.1 \le \ohsq \le 0.3$ in this `generic' domain requires
moderate values $m_{1/2} \lappeq 400$~GeV and $m_0 \lappeq 200$~GeV, where
the LHC may be able to make detailed measurements of the sparticle
spectrum and hence the CMSSM parameters~\cite{Paige}. We return later to a
more careful consideration of the individual $\Delta^\Omega_i$ and the
uncertainties in the corresponding $a_i$.

It is apparent in panel (a) of Fig.~\ref{fig:overall} that the overall
sensitivity $\Delta^\Omega$ increases at both large and small values of
$m_0 / m_{1/2}$. The increase in $\Delta^\Omega$ at large $m_0 / m_{1/2}$
is primarily due to the approach to the direct-channel $\chi \chi
\rightarrow h$ pole. The enhanced annihilation cross section reduces the
relic density to an acceptable level for finely tuned values of $m_{1/2}$,
which is the reason $\Delta^\Omega$ takes on values in excess of 100
there.  However, a close approach to this pole is forbidden by the LEP
lower limits on the chargino mass $m_{\chi^\pm}$, and is also disfavoured
by the LEP lower limit $m_h > 113.5$~GeV~\cite{LEPHiggs}, making this
point somewhat moot. 

The increase in $\Delta^\Omega$ close to a ray in the $(m_{1/2}, m_0)$
plane at small $m_0 / m_{1/2}$ is due to the importance of
coannihilation~\cite{EFOSi}, whose significance varies with $m_{\tilde
\ell} - m_\chi$ and hence the CMSSM parameters. However, we still find
that $\Delta^\Omega < 20$ in this coannihilation region, so the relic
density does not require excessive fine-tuning in order to fall within the
preferred range $0.1 \le \ohsq \le 0.3$. On the other hand, the LHC may
not be able to provide very detailed measurements of the sparticle
spectrum in this region~\cite{CMS}, so it may not facilitate a very
accurate calculation of $\ohsq$. On the bright side, we note that this
region does not agree well~\cite{ENO} with the value of $g_\mu - 2$
reported recently~\cite{g-2}. 

We do not show $(m_{1/2}, m_0)$ planes for other low-to-moderate values of
$\tan \beta \lappeq 25$, but simply remark that they are qualitatively
similar to Fig.~\ref{fig:overall}(a) for both signs of $\mu$. In
particular, there are qualitatively similar zones where $\Delta^\Omega
\lappeq 10$, or even $\lappeq 3$. These regions are also generally
compatible with $g_\mu - 2$~\cite{g-2}. However, it should be remembered
that the $b \rightarrow s \gamma$ constraint~\cite{bsg} (not shown here) 
excludes domains of small $m_{1/2}$ which increase as $\tan \beta$
increases, and are larger for $\mu < 0$. 

Panel (b) of Fig.~\ref{fig:overall} displays the $(m_{1/2}, m_0)$ plane
for $\mu <0$ and $\tan \beta = 35$, near the upper limit for which we find
extensive regions of acceptable electroweak vacua for this sign of $\mu$
and our default choices of $m_t$ and $m_b$~\cite{EFGOSi}. We note that the
sensitivity $\Delta^\Omega$ is generally higher than in panel (a) for
$\tan \beta = 10$, foreshadowing the breakdown of the electroweak vacuum
conditions. We also see a `funnel' at $m_0 \sim m_{1/2}$, where the relic
density varies rapidly, reflecting the importance of direct-channel $\chi
\chi \rightarrow H,A$ pole annihilations, so that $\Delta^\Omega$ is
large. Indeed, $\Delta^\Omega \sim 100$ in the cosmological funnel, and
even exceeds 1000 deep in the pole region where the relic density is very
small.  The sensitivity measure $\Delta^\Omega$ is significantly larger
than for $\tan \beta = 10$ also at larger values of $m_0 / m_{1/2} \sim
2$, reflecting the fact that the preferred range of $m_0$ increases
relatively rapidly as $\tan \beta$ increases and the rapid-annihilation
`funnel' moves to higher $m_0 \sim m_{1/2}$. The behaviour of
$\Delta^\Omega$ in the coannihilation region of Fig.~\ref{fig:overall}(b) 
is qualitatively similar to that in Fig.~\ref{fig:overall}(a), whilst
being somewhat more elevated. In the good cosmological region with low
$m_{1/2}$, $\Delta^\Omega \ga 20$.

Panel (c) of Fig.~\ref{fig:overall} displays the case $\mu > 0$ and
$\tan\beta = 50$, which is again close to the upper limit for which we
find extensive regions of acceptable electroweak vacua for this sign of
$\mu$ and our default choices of $m_t$ and $m_b$~\cite{EFGOSi}. Panel (c) 
has many qualitative features in common with panel (b), notably the very
elevated values of $\Delta^\Omega$ around a rapid-annihilation `funnel',
and the somewhat elevated values of $\Delta^\Omega$ in the regions at
higher and lower values of $m_0 / m_{1/2}$.

Finally, panel (d) of Fig.~\ref{fig:overall} displays another case with
$\mu > 0$ and $\tan\beta = 10$, this time for $m_t = 171$~GeV. Its
features are rather similar to those of panel (a) for $m_0 \lappeq
800$~GeV, but now we also see the `focus-point' region of acceptable
$\ohsq$ for $m_0 \gappeq 1000$~GeV~\footnote{At higher values of $m_t$, we
find the focus-point region at higher values of $m_0$.}.  The
`focus-point' region adjoins the (mauve) shaded region where we do not
find a consistent electroweak vacuum. The fact that the `focus-point'
region moves rapidly with a small change in $m_t$ largely explains the
high values of the sensitivity parameter $\Delta^\Omega \sim 500$ in this
region: analogous high sensitivity to $m_t$ can be seen in Fig.~9 of the
second paper in~\cite{FM}~\footnote{We thank  K.~Matchev for
discussions on this point.}.

\begin{figure}
\vspace*{-0.75in}
\hspace*{-.10in}
\begin{minipage}{8in}
\epsfig{file=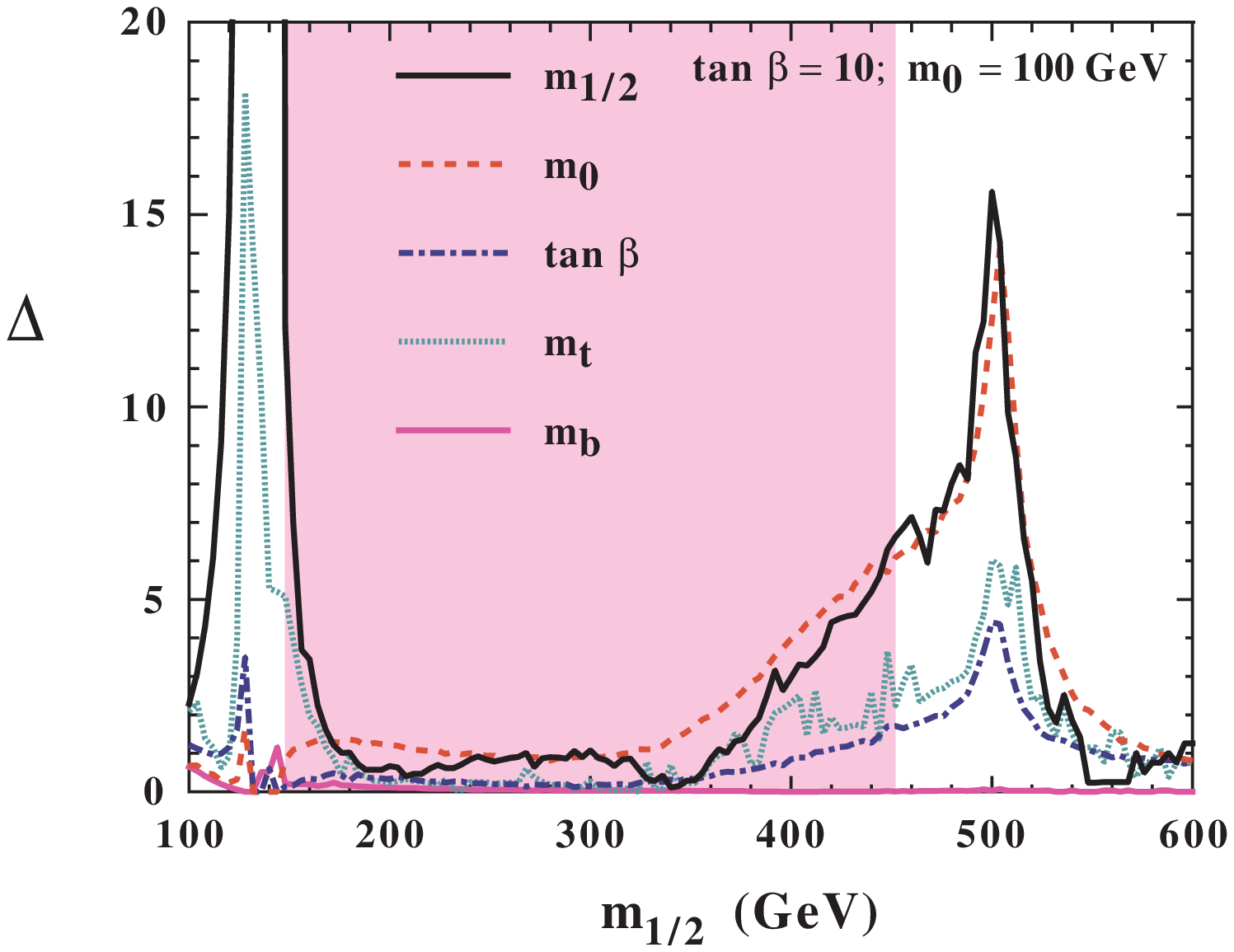,height=2.5in}
\epsfig{file=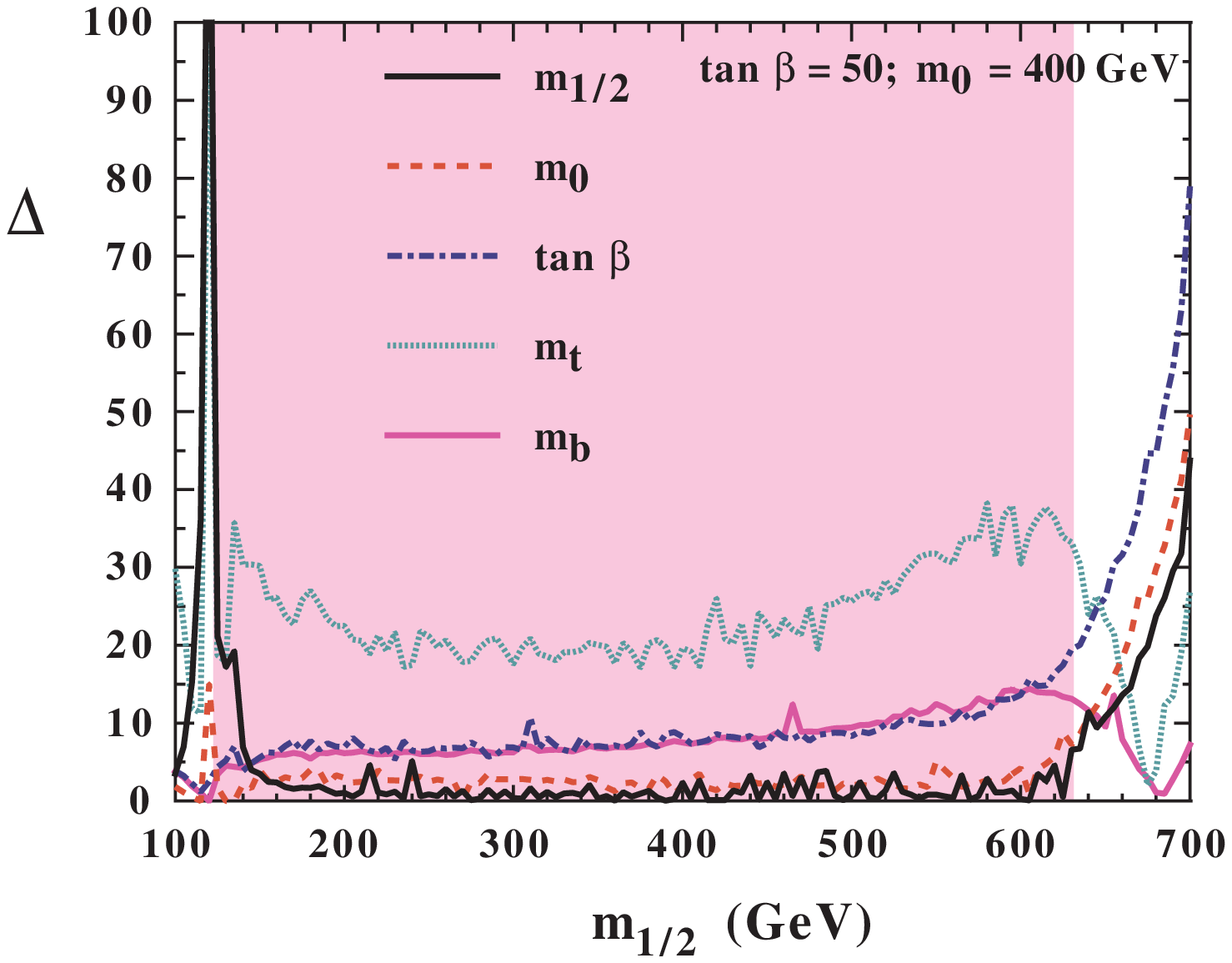,height=2.5in} \hfill
\end{minipage}
\begin{minipage}{8in}
\epsfig{file=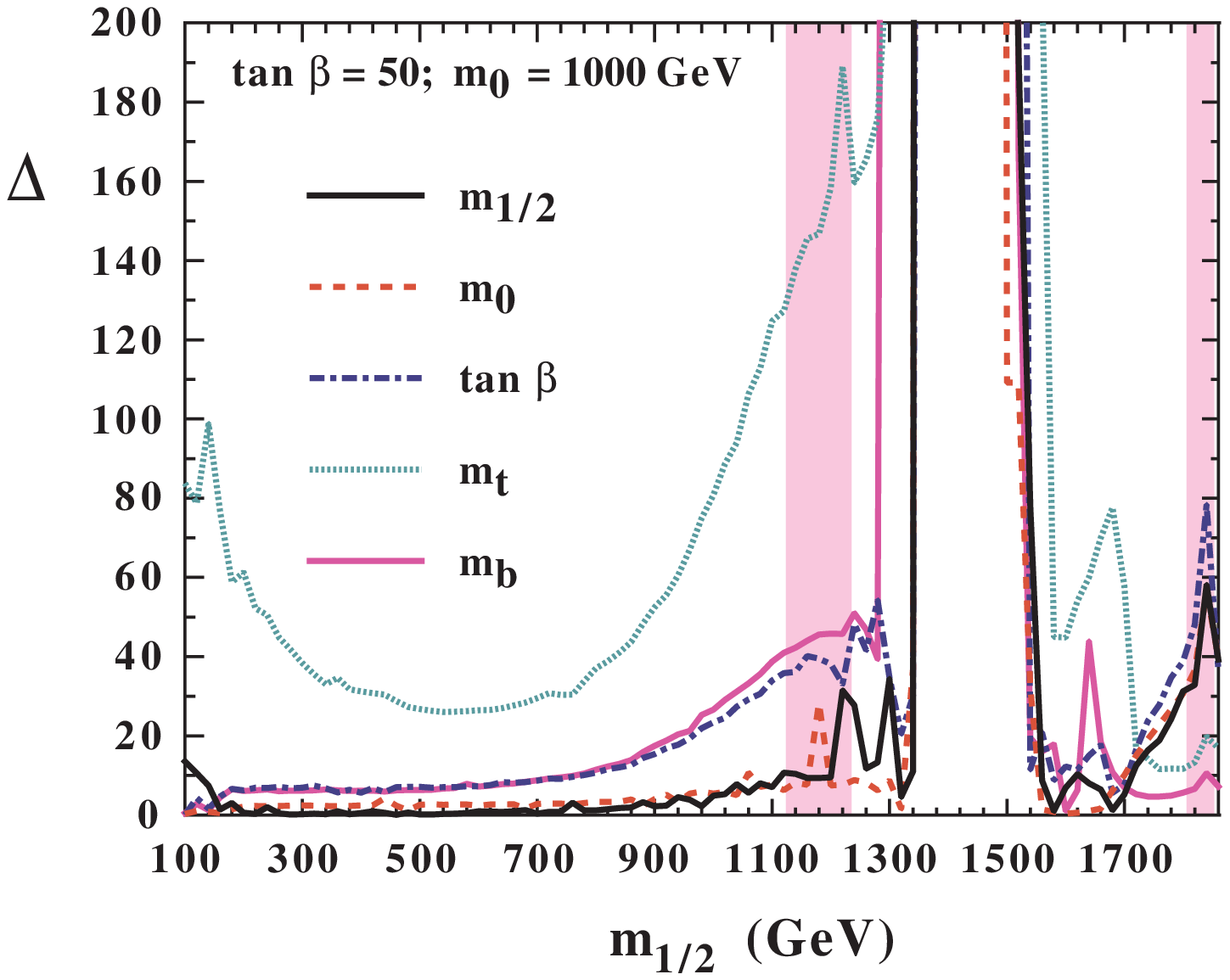,height=2.5in}
\epsfig{file=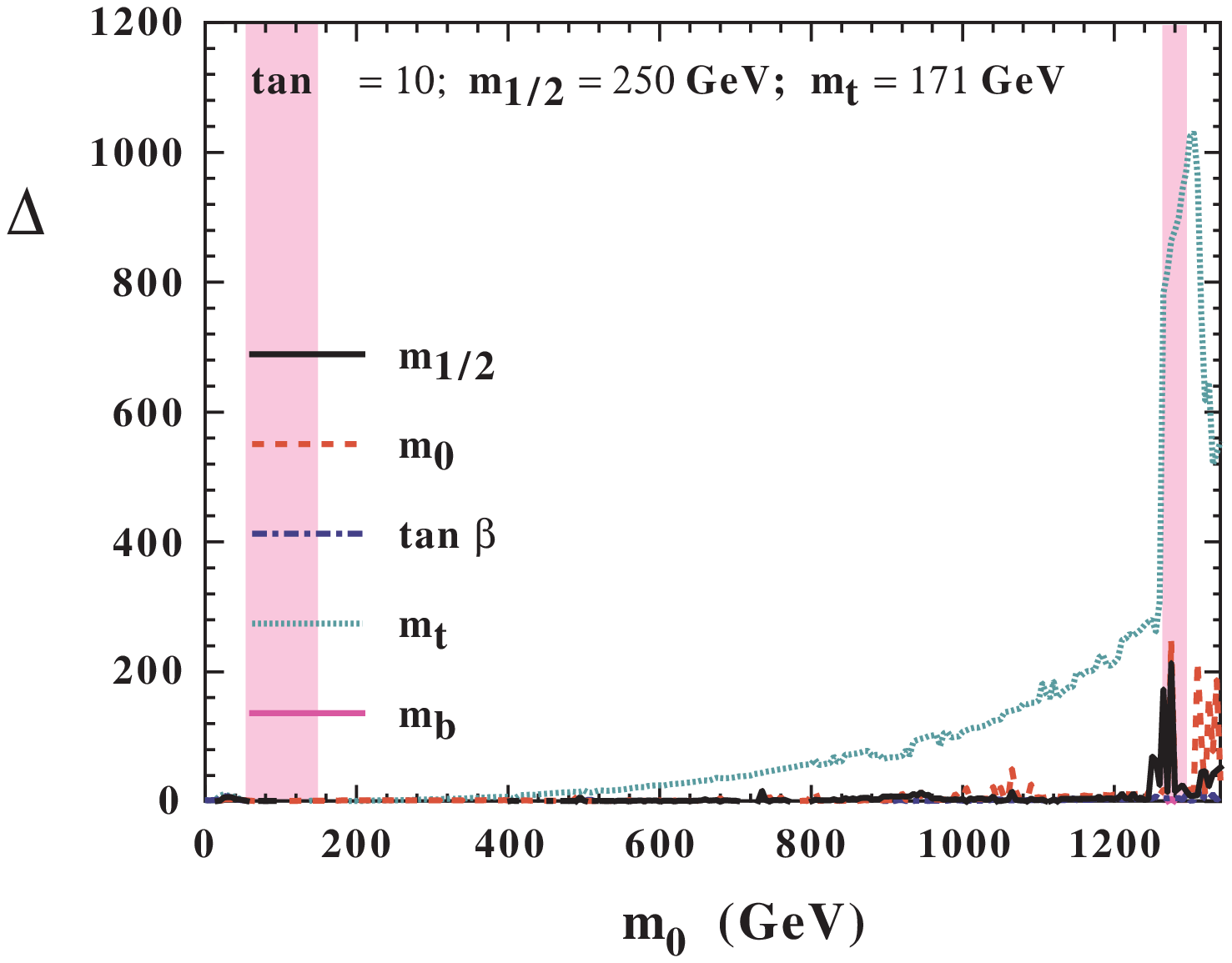,height=2.5in} \hfill
\end{minipage}
\caption{\label{fig:slices}
{\it Fine-tuning sensitivities along
illustrative slices through the CMSSM parameter space: for (a) $m_0 =
100$~GeV when $\tan \beta = 10, \mu > 0, m_t = 175$~GeV, (b) $m_0
= 400$~GeV when $\tan \beta = 50, \mu > 0, m_t = 175$~GeV, (c)
$m_0 = 1000$~GeV when $\tan \beta = 50, \mu > 0, m_t = 175$~GeV,
and (d) for $m_{1/2} = 250$~GeV when $\tan \beta = 10, \mu > 0,
m_t = 171$~GeV, all for $A_0 = 0$. The sensitivities $\Delta^\Omega_i$ to
different input
parameters are plotted using different line styles, as indicated. The
sensitivity to $A_0$ is always negligible, and is not shown. The vertical
(pink) shaded bands have $0.1 \le \ohsq \le 0.3$.
}}
\end{figure}

As an aid to better understanding of the origins of the variations in the
overall sensitivity measure $\Delta^\Omega$ in Fig.~\ref{fig:overall}, we
display in Fig.~\ref{fig:slices} the values of all the individual
$\Delta^\Omega_i$ along various illustrative slices through the CMSSM
parameter space at constant $m_0$ or $m_{1/2}$. The vertical (pink) shaded
strips in the panels of Fig.~\ref{fig:slices} show the regions where the
relic density falls within the preferred range $0.1 \le \ohsq \le 0.3$.

Panel (a) of Fig.~\ref{fig:slices} is for $m_0 = 100$~GeV when $\tan \beta
= 10, \mu > 0, A_0 = 0, m_t = 175$~GeV and $m_b(m_b)^{\overline {MS}} =
4.25$~GeV, corresponding to a slice of Fig. \ref{fig:overall}(a). Looking
first at the `generic' region where $150$~GeV $\lappeq m_{1/2} \lappeq
400$~GeV, we see that the dominant sensitivities $\Delta^\Omega_i$ are
those to $m_0$ and $m_{1/2}$, both of which are close to unity. Next in
importance is the sensitivity to $m_t$, which is ${\cal O}(1/2)$. Finally,
the sensitivities to $\tan \beta, m_b(m_b)^{\overline {MS}}$ and
particularly $A_0$ are rather negligible in this domain. Clearly visible
is a sharp increase in some of the $\Delta^\Omega_i$ for $m_{1/2} \lappeq
150$~GeV, dominated by jumps in the sensitivity to $m_{1/2}$ and $m_t$ as
$m_\chi$ traverses the value $m_h/2$, and the rate of $\chi \chi
\rightarrow h$ annihilation changes rapidly. However, as already
mentioned, this pole region is not relevant to the dark matter issue,
because it is excluded by the LEP constraint on $m_{\chi^\pm}$ and would,
in any case, give a very suppressed relic density $\ohsq \ll 0.1$ in all
but a very narrow strip in $m_{1/2}$. Also visible for $m_{1/2} \gappeq
400$~GeV are more gradual rises in some of the $\Delta^\Omega_i$ as
$m_\chi - m_{\tilde \tau_1} \rightarrow 0^-$ in the coannihilation
region~\cite{EFOSi}, followed by falls in the disallowed domain where
$m_\chi > m_{\tilde \tau_1}$. The sensitivities to $m_0$ and $m_{1/2}$ are
the largest, and are very similar, as is to be expected because they are
the key parameters controlling $m_\chi - m_{\tilde \tau_1}$. This behavior
is easily understood when one takes into account the exponential
sensitivity of coannihilation to $m_\chi - m_{\tilde \tau_1}$ and
recongizes the strong dependence of $m_{\tilde \tau_1}$ on both $m_{1/2}$
and $m_0$ and $m_\chi$ on $m_{1/2}$. The sensitivity to
$m_t$ arises from the renormalization-group equations used to determine
the low-energy parameters in terms of the GUT-scale input parameters, and
$\tan \beta$ also enters in the determination of the $m_\chi - m_{\tilde
\tau_1} = 0$ contour. We see that, even in this region, none of the individual
$\Delta^\Omega_i$ exceeds 15, and recall that, as seen in
Fig.~\ref{fig:overall}(a), the combined $\Delta^\Omega \lappeq 20$ in the
region of preferred relic density $0.1 \le \ohsq \le 0.3$ for $\tan \beta
= 10$. 

Panel (b) of Fig.~\ref{fig:slices} shows a slice at fixed $m_0 = 400$~GeV
through Fig.~\ref{fig:overall}(c), for $\tan \beta = 50, \mu > 0,
A_0 = 0, m_t = 175$~GeV and $m_b(m_b)^{\overline {MS}} = 4.25$~GeV.  In
the generic region where $m_{1/2} \lappeq 600$~GeV, which includes the
range where $0.1 \le \ohsq \le 0.3$, we see that the dominant sensitivity
(between 20 and 30) is that to $m_t$, which is associated with the
`squeezing' of this region as the `funnel' moves towards the vertical
axis. Next in importance for $m_{1/2} \lappeq 600$~GeV are the
sensitivities ($\lappeq 10$) to $\tan \beta$ and $m_b(m_b)^{\overline
{MS}} = 4.25$~GeV, which also originate from their effects on the
`funnel'. The sensitivities to $m_0$ and $m_{1/2}$ are much smaller, even
comparable to those in panel (a), and the sensitivity to $A_0$ is
negligible, as always for $A_0 = 0$. The region of high sensitivity when
$m_{1/2} \lappeq 150$~GeV is actually excluded by the $b \rightarrow s
\gamma$ constraint, which imposes $m_{1/2} \gappeq 300$~GeV for this value
of $m_0$ and the other parameters.

Panel (c) of Fig.~\ref{fig:slices} shows a slice at fixed $m_0 =
1000$~GeV, again through Fig.~\ref{fig:overall}(c), for $\tan \beta = 50,
\mu > 0, A_0 = 0, m_t = 175$~GeV and $m_b(m_b)^{\overline {MS}} =
4.25$~GeV.  As also seen in Fig.~\ref{fig:overall}(c), there are much
narrower ranges of $m_{1/2}$ where $0.1 \le \ohsq \le 0.3$: one on the
left side of the rapid annihilation `funnel', a much narrower region on
the right side when $m_{1/2} \sim 1600$~GeV, that is not shown in
Fig.~\ref{fig:slices}(c), and another narrow region in the coannihilation
region when $m_{1/2} \sim 1800$~GeV. The dominant sensitivity to the left
of the `funnel' is that to $m_t$, followed by those to $\tan \beta$ and
$m_b$. These all increase as the `funnel' is approached, reflecting its
sensitivities to these parameters. In the coannihilation region of
Fig.~\ref{fig:slices}(c), the sensitivities to $\tan \beta, m_0$ and
$m_{1/2}$ dominate as $m_{\tilde \tau_1} - m_\chi \rightarrow 0^+$, and
are all more important than for $\tan \beta = 10$, as shown in
Fig.~\ref{fig:slices}(a). 

Finally, panel (d) of Fig.~\ref{fig:slices} shows a slice at fixed
$m_{1/2} = 250$~GeV through Fig.~\ref{fig:overall}(d), for $\tan \beta =
10, \mu > 0, A_0 = 0, m_t = 171$~GeV and $m_b(m_b)^{\overline {MS}} =
4.25$~GeV. This cuts through both the `focus-point' region and the
`generic' domain, for which a different slice was shown in panel (a).  The
most noticeable feature is a strong growth in the sensitivity to $m_t$ as
the `focus-point' region is approached, with a maximum value
$\Delta^\Omega_{m_t} \simeq 1000$: see also Fig.~9 of the second paper
in~\cite{FM}. We also note increased sensitivities to $m_{1/2}$ and $m_0$
in this region: $\Delta^\Omega_{m_{1/2}, m_0} \simeq 200$, reflecting the
narrowness of the `focus-point' strip in the $(m_{1/2}, m_0)$ plane. The
sensitivities in the `generic' domain at smaller $m_{1/2}$ are invisible
in this plot, but are very similar to those shown in
Fig.~\ref{fig:slices}(a), namely $\Delta^\Omega_i \lappeq {\cal O}(1)$.

Finally, we consider what light this analysis casts on the accuracy with
which LHC measurements might eventually enable $\ohsq$ to be
calculated~\cite{Drees}.  We assume that $\delta m_t / m_t = 1 \%$ in the
LHC era, and that $\delta m_b / m_b = 5 \%$~\cite{mb} in all cases.
Detailed studies of the precision with which a combination of LHC
measurements could constrain CMSSM parameters have been made for a limited
number of benchmark points~\cite{Paige,PPRS,fit2,fit3}.  Unfortunately,
these LHC benchmark points are now outdated, e.g., because the relic
density is too high or because $m_h$ is too low, and they are often bad
also for $g_\mu - 2$ and/or $b \rightarrow s \gamma$. However, we select
for our analysis two LHC points that yield $\ohsq < 0.3$, and attempt to
extract from them useful indicators for points that yield $\ohsq$ in the
preferred range. 

{\bf LHC Point 5}: This is the LHC point for which the most detailed
studies are available~\cite{Paige,PPRS,fit2}. It has $\mu > 0$ and the
following values
of the CMSSM parameters~\footnote{The values of $A_0$ for this and the
other LHC points are essentially irrelevant, because
$\Delta^\Omega_{A_0} \ll 1$, and we set $A_0 = 0$ in the following.}:
\begin{equation}
m_0 = 100~{\rm GeV}, \; m_{1/2} = 300~{\rm GeV}, \; \tan \beta = 2
\label{LHC5}
\end{equation}
corresponding, according to our calculations, to $\ohsq = 0.22$ (within
the preferred range) and $m_h \simeq 91$~GeV~\cite{FH} (which is
excluded by LEP). Moreover, though its value of $b \rightarrow s \gamma$ 
is satisfactory, its value of $g_\mu - 2$ is too small. However, it may
serve as a useful indicator. At this point, a number of spectroscopic
measurements would have been possible at the LHC~\cite{Paige,PPRS,fit2},
and
the errors in the LHC determinations of the numerical parameters were
estimated to be: 
\begin{equation}
\delta m_0 = 3.6~{\rm GeV}, \; \delta m_{1/2} = 5.0~{\rm GeV}, \; \delta
\tan \beta = 0.18.
\label{errors5}
\end{equation}
Extending our analysis of the $\Delta^\Omega_i$ to
this specific extra case, we find the following sensitivities to
parameters: 
\begin{equation}
\Delta^\Omega_{m_0} = 0.80, \; \Delta^\Omega_{m_{1/2}} = 0.92, \;
\Delta^\Omega_{\tan \beta} = 0.07, \; \Delta^\Omega_{m_t} \simeq
\Delta^\Omega_{m_b} \simeq 0.
\label{sensitivity5}
\end{equation}
Combining in quadrature the errors in (\ref{errors5}) with the
sensitivities (\ref{sensitivity5}) in the calculation of
$\ohsq$, we estimate
\begin{equation}
{\delta \ohsq \over \ohsq} \; \gappeq \; {1 \over 30};
\label{result5}
\end{equation}
where the inequality sign recalls that there are certainly other errors in
the calculation of $\ohsq$, that may not be negligible. However, we infer
from (\ref{result5}) that an accurate calculation of $\ohsq$ may be
possible in `generic' domains of the allowed CMSSM parameter space for
moderate $\tan \beta$.

{\bf LHC Point 6}: This~\cite{fit3} is the only LHC point with large
$\tan \beta$. It has $\mu > 0$ and the following
values of the CMSSM parameters:
\begin{equation}
m_0 = 200~{\rm GeV}, \; m_{1/2} = 200~{\rm GeV}, \; \tan \beta = 45
\label{LHC6}
\end{equation}
corresponding, according to our calculations, to $\ohsq = 0.045$ (below 
our preferred range, but not excluded) and $m_h
= 112$~GeV (which may be allowed by LEP when one allows for theoretical
uncertainties). However, neither $b \rightarrow s \gamma$ nor $g_\mu
-2$ are satisfactory for this point. The errors in the LHC   
determinations of the CMSSM parameters were estimated to be:
\begin{equation}
\delta m_0 = 29~{\rm GeV}, \; \delta m_{1/2} = 9~{\rm GeV}, \; \delta \tan
\beta = 5.
\label{errors6}
\end{equation}
In this case, we find that the parameter sensitivities are somehwat more
elevated:
\begin{equation}
\Delta^\Omega_{m_0} = 2.45, \; \Delta^\Omega_{m_{1/2}} = 1.65, \;
\Delta^\Omega_{\tan \beta} = - 5.18, \; \Delta^\Omega_{m_t} = 9.0, \;
\Delta^\Omega_{m_b} = - 3.51,
\label{sensitivity6}
\end{equation}
in view of which we conclude that
\begin{equation}
{\delta \ohsq \over \ohsq} \; = \; {\cal O}(1)
\label{result6}
\end{equation}
in this case. 

We note, moreover, that, for this value of $\tan \beta$, $\ohsq$ is large
enough to be in the range preferred by cosmology only if larger values of
$m_0$ and/or $m_{1/2} \gappeq 400$~GeV are chosen.  We recall that LHC
Points 1 and 2 had $m_0 = m_{1/2} = 400$~GeV~\cite{Paige}, and that in
these cases the limited LHC measurements did not provide any accuracy in
the determination of $m_0$. (These points also had $\ohsq \gappeq 1.6$,
acceptable $b \rightarrow s \gamma$ and unacceptable $g_\mu - 2$.) We
conclude from this discussion and (\ref{result6})  that an accurate
calculation of $\ohsq$ may not be possible at large $\tan \beta$ using LHC
data alone. 

For the record, we recall that LHC Point 3~\cite{Paige} had $m_0 =
200~{\rm GeV}, m_{1/2} = 100~{\rm GeV}$, $\tan \beta = 2$ and $\mu < 0$,
leading to $m_h \sim 67$~GeV~\cite{FH}, which is far too small.  This
points also had $\ohsq \simeq 0.38$ (rather too high) and unacceptable
$g_\mu - 2$, though $b \rightarrow s \gamma$ was satisfactory. We do not
discuss this point in detail, but note that, like at Point 5, $\ohsq$
could in principle be calculated quite accurately using LHC data. Finally,
LHC Point 4 has $m_0 = 800~{\rm GeV}, m_{1/2} = 200~{\rm GeV}, \tan \beta
= 10$, leading to $\ohsq = 5.6$, rendering it uninteresting for this
analysis. For completeness, we note that this point had $m_h
\simeq 111$~GeV, acceptable $b \rightarrow s \gamma$ and unacceptable
$g_\mu - 2$. We also note that, although some sparticle measurements are
possible in the coannihilation region~\cite{CMS}, it seems unlikely that
LHC measurements alone will constrain the CMSSM parameters sufficiently to
enable $\ohsq$ to be calculated accurately. 

To conclude: We have demonstrated in this paper that there are `generic'
domains of CMSSM parameter space at moderate $\tan \beta$ where the
sensitivity $\Delta^\Omega$ of the relic density $\ohsq$ is rather small. 
Thus, obtaining $\ohsq$ in the range preferred by astrophysics and
cosmology does not require `fine-tuning' of the values of the CMSSM
parameters. The sensitivity of $\ohsq$ to the CMSSM parameters is somewhat
increased in the coannihilation region~\cite{EFOSi}, but not to an
alarming extent. It is also increased at large $\tan \beta$, particularly
in the `funnel' regions where rapid $\chi \chi \rightarrow H, A$
annihilations are important~\cite{EFGOSi}. We also found large values of
$\Delta^\Omega$ in the `focus-point' region~\cite{FM}, where the CMSSM
parameters and particularly $m_t$ must be adjusted for a given set of
supersymmetric input parameters, if $\ohsq$ is to fall within the
preferred range. The tracking of the individual sensitivities,
$\Delta^\Omega_i$ clarifies which parameters must be measured and treated
carefully in order to calculate $\Omega h^2$ reliably. 

In the generic regions with low $\Delta^\Omega$, LHC
measurements~\cite{Paige} may enable $\ohsq$ to be calculated accurately. 
It would be interesting to study how accurately the CMSSM parameters could
be measured at a new set of benchmark points that respect the constraints
imposed by LEP and other recent experiments~\cite{new}, both at the LHC
and with a
possible linear $e^+ e^-$ collider. As already mentioned, there are
clearly cases where the LHC alone cannot determine the CMSSM parameters
with sufficient precision to enable $\ohsq$ to be calculated accurately,
and it would be interesting to see how a linear $e^+ e^-$ collider could
contribute. A successful, accurate calculation of $\ohsq$ on the basis of
accelerator data would surely be the culmination of supersymmetric dark
matter studies, making this a worthwhile objective to pursue. 

\vskip 0.5in
\vbox{
\noindent{ {\bf Acknowledgments} } \\
\noindent  
The work of K.A.O. was partially supported by DOE grant
DE--FG02--94ER--40823.}

\end{document}